\documentclass[runningheads]{llncs}
\usepackage[T1]{fontenc}
\usepackage{amsmath,amssymb,amsfonts}
\usepackage{float}
\usepackage{orcidlink}
\usepackage{subcaption}
\usepackage{algorithm}
\usepackage{algorithmic}
\usepackage{graphicx}
\usepackage{textcomp}
\usepackage{xcolor}
\usepackage{todonotes}
\usepackage{listings}
\usepackage{adjustbox}
\usepackage{dirtytalk}

\definecolor{jsonkey}{rgb}{0.4, 0.1, 0.1}
\definecolor{jsonvalue}{rgb}{0.1, 0.1, 0.6}
\definecolor{jsonstring}{rgb}{0.1, 0.6, 0.1}
\definecolor{pythonkeyword}{rgb}{0.13, 0.29, 0.53}
\definecolor{pythonstring}{rgb}{0.31, 0.60, 0.02}
\definecolor{commentcolor}{rgb}{0.56, 0.35, 0.01}
\definecolor{backgroundcolor}{rgb}{0.95, 0.95, 0.92}

\lstdefinestyle{pythonstyle}{
    language=Python,
    basicstyle=\ttfamily\small,
    keywordstyle=\color{pythonkeyword}\bfseries,
    stringstyle=\color{pythonstring},
    commentstyle=\color{commentcolor}\itshape,
    showspaces=false,
    showstringspaces=false,
    showtabs=false,
    tabsize=2,
    breaklines=true,
    breakatwhitespace=true,
    rulecolor=\color{black},
    numbersep=10pt,
    framexleftmargin=20pt,
    framextopmargin=5pt,
    framexbottommargin=5pt,
    escapeinside={(*@}{@*)},
}

\begin{document}

\title{NLP-Guided Synthesis: Transitioning from Sequential Programs to Distributed Programs}
\titlerunning{NLP-Guided Synthesis} 

\author{Arun Sanjel~\orcidlink{0009-0009-2453-8538},
Bikram Khanal~\orcidlink{0000-0003-2292-520X},
Greg Speegle~\orcidlink{0000-0003-1739-0271},
Pablo Rivas~\orcidlink{0000-0002-8690-0987}}
\authorrunning{A. Sanjel et al.}
\institute{Department of Computer Science, Baylor University, Texas, USA
\email{\{arun\_sanjel1,bikram\_khanal1,greg\_speegle,pablo\_rivas\}@baylor.edu}}

\maketitle

\begin{abstract}
As the need for large-scale data processing grows, distributed programming frameworks like PySpark have become increasingly popular. However, the task of converting traditional, sequential code to distributed code remains a significant hurdle, often requiring specialized knowledge and substantial time investment. While existing tools have made strides in automating this conversion, they often fall short in terms of speed, flexibility, and overall applicability. In this paper, we introduce ROOP, a groundbreaking tool designed to address these challenges. Utilizing a BERT-based Natural Language Processing (NLP) model, ROOP automates the translation of Python code to its PySpark equivalent, offering a streamlined solution for leveraging distributed computing resources. We evaluated ROOP using a diverse set of 14 Python programs comprising 26 loop fragments. Our results are promising: ROOP achieved a near-perfect translation accuracy rate, successfully converting 25 out of the 26 loop fragments. Notably, for simpler operations, ROOP demonstrated remarkable efficiency, completing translations in as little as 44 seconds.  Moreover, ROOP incorporates a built-in testing mechanism to ensure the functional equivalence of the original and translated code, adding an extra layer of reliability. This research opens up new avenues for automating the transition from sequential to distributed programming, making the process more accessible and efficient for developers.
\end{abstract}

\section{Introduction}
In recent years, distributed programming has gained significant traction, driven by advancements in tools that facilitate large-scale data cluster computations and the growing necessity for analyzing extensive datasets. However, the level of parallelism that can be achieved is often constrained by the limited range of operations supported by existing parallel programming languages. Popular Data Intensive Scalable Computing (DISC) frameworks such as MapReduce \cite{dean2008mapreduce}, Flink \cite{flink}, Spark \cite{zaharia2010spark}, and PySpark \cite{pyspark} simplify these complex details and provide highly efficient implementations. However, because each framework's underlying architecture is unique, application development on these systems may take much longer and require more labor than building sequential applications in languages such as Java or Python. As a consequence, productivity would be greatly increased by automatically turning a sequential program into a distributed program in a suitable framework.

Our previous work, \textit{Tyro} \cite{Tyro} and its successor \textit{Tyro v2} \cite{tyrov2} were able to transform different sequential Python programs into semantically equivalent PySpark applications. Tyro converts sequential code pieces to their corresponding distributed operations by utilizing an Abstract Syntax Tree (AST), static program analysis, Gradual Program Synthesis (GRASSP) \cite{grassp}, and unit testing \cite{unit_test}. Tyro v2 leverages a graph-based loop extraction method to capture additional sequential processes,  enhancing component extraction where each component is a piece of code. Tyro's intuitive approach begins by searching for the simplest operation that might be equivalent to a specific code fragment and then gradually enhances operations used in the synthesis process. Tyro divides the search space into pieces and begins with a tiny domain of distributed operations rather than considering a vast solution space at one time. Although utilizing the GRASSP method reduces the search space for each iteration, it still has to  enumerate a large space. As such, Tyro uses a hand-crafted algorithm to improve the  search process. This algorithm is a bottleneck in terms of generalizing Tyro to handle larger and more complex programs.

However, neural models provide an alternative for program synthesis that does not require a human-crafted search strategy \cite{abolafia2018neural,hong2021latent,kalyan2018neuralguided}. New neural synthesis systems for program synthesis have emerged in recent years, many of which employ similar concepts such as neural machine translations \cite{grammarandrl}. Notably, large language models like GPT~\cite{GPT} and Codex~\cite{chen2021evaluating} have demonstrated the potential of using NLP for tasks closely related to code translation and synthesis. These developments suggest that NLP techniques could offer a more scalable and generalizable solution for automating the transition from sequential to distributed code.

In this context, we present \textit{ROOP}, a novel tool that leverages a NLP-based Bidirectional Encoder Representations from Transformers (BERT) model for code translation. ROOP takes a sequential Python program that operates iteratively on a dataset, along with accompanying test cases as input. It first identifies \say{Loop Fragments} (a term we introduce to describe iterative statements or code snippets) within the code. These Loop Fragments are then processed by our BERT model, which predicts their corresponding PySpark API calls. Following this, the predicted API calls are refactored according to a predefined set of rules and structures. The refactored code is then subjected to the provided test cases, serving as a program equivalence verifier. Once the verifier successfully passes these test cases, ROOP generates a complete, executable PySpark program. This approach not only enables a more accurate translation but also incorporates unit testing as a crucial step for ensuring code equivalence, thereby aligning closely with the true purpose of code generation.
Our key contributions are:
\begin{itemize}
    \item We propose ROOP, a tool that uses a BERT-based NLP model, innovatively synthesizing distributed PySpark code from a given Python code and validation processes to ensure both accuracy and functional equivalence.
    \item  We  evaluate ROOP's performance across a diverse range of Python programs, showing high translation accuracy and quick translation.
 \end{itemize}

The rest of this paper covers several key areas. First, we look at existing research to set the stage for our own work. Next, we go into the technical details of our tool, ROOP, and how it operates. We also share the results of tests we've run to show that our approach works well. Finally, we discuss what these results mean and what could be explored in future research, wrapping up a summary of what we've achieved in making it easier to convert Python code into PySpark applications.

\section{Related Work}

The realm of program synthesis, notably the transition from sequential to distributed programming, has been an active area of research over the past few years~\cite{casper,sanjel2020tyro,mold,diablo}. The complexities inherent in translating code structures, combined with the need to maintain the semantics of the original code, present unique challenges that have spurred numerous methodologies and tools. This section provides an overview of prior research in this domain, setting the context for our NLP-driven approach
\subsection{Sequential to Distributed Translation}
The shift from sequential to distributed programming paradigms is driven by the increasing demands of big data processing. While sequential programs are typically easier to write and debug, they often lack the scalability required for large-scale data processing tasks. Distributed frameworks, on the other hand, provide scalability but introduce complexities in terms of data partitioning, task distribution, and fault tolerance.

Several tools and methodologies have been proposed to assist developers in this transition. For instance, Casper~\cite{casper} stands out as a significant contributor. Casper uses search algorithms to convert sequential Java code into the MapReduce framework. Casper employs program synthesis to generate program summaries, which are then verified for semantic accuracy using a theorem prover. While effective, its search-based methodology has limitations in handling a broad range of coding complexities.

On a similar note, Mold~\cite{mold} offers an alternative approach, targeting the Apache Spark runtime. Mold converts input code into a functional form, guided by rewrite rules to find an optimal MapReduce implementation. It introduces a unique technique for handling irregular loop dependencies, enhancing parallelism. In evaluations, Mold has proven effective even for codes with complex updates. 

Following the same, the authors in \cite{diablo} introduce a novel framework  DIABLO designed to translate programs expressed as array-based loops into distributed data-parallel programs. Targeting scientists who are well-versed in numerical analysis but less familiar with Big Data analytics, this framework aims to simplify the transition from loop-based to data-parallel programming. Built on top of Spark, the prototype has demonstrated both greater generality and efficiency compared to existing solutions, even when benchmarked against hand-written programs.

Unlike Casper and Mold, which focus on MapReduce, our tool ROOP extends its reach to include more operations in PySpark. While DIABLO also uses Spark, it's not as flexible as ROOP. Our tool uses data-driven synthesis, making it easier to adapt to different frameworks and various distributed operations.  This makes ROOP a more flexible and robust solution for automating the translation of data-intensive applications, catering to a wider array of programming paradigms and data processing requirements

\subsection{NLP in Code Translation}
The application of NLP to code-related tasks has opened up a plethora of research avenues, each marked by innovative techniques and promising results. One such significant venture is semantic parsing, which aims to convert natural language queries into executable code snippets. Youssef et al. spearheaded efforts in this domain, leveraging Artificial Neural Networks (ANN)~\cite{jain1996artificial} and Long Short-Term Memory (LSTM)~\cite{hochreiter1997long} based Recurrent Neural Networks (RNN)~\cite{amari1972learning} to generate Python code from natural language descriptions~\cite{code_to_text}.

Complementing this, the realm of binary code analysis has also benefited from NLP methodologies. Zuo et al. adapted techniques originally developed for NLP to perform large-scale binary code similarity assessments. Their work serves as a cornerstone for exploring further research in code translation and similarity evaluation~\cite{zuo2018neural}. Structured prediction, a concept deeply rooted in NLP, has found relevance in code generation as well. Wang et al. introduced CODE4STRUCT~\cite{wang2023code4struct}, a specialized model trained on a hybrid dataset of text and code. This model employs text-to-structure translation mechanisms to accomplish structured prediction tasks, thereby creating a symbiotic relationship between natural language and code structures.

In conclusion, the achievements made in semantic parsing, binary code analysis, structured prediction, and transfer learning, along with the high performance exhibited by large language models, underline the feasibility and value of pursuing data-driven or NLP-guided synthesis for complex tasks such as sequential to distributed code translation.

\section{System Architecture}
\begin{figure}[t!]
    \centering
    \includegraphics[width=0.8\textwidth]{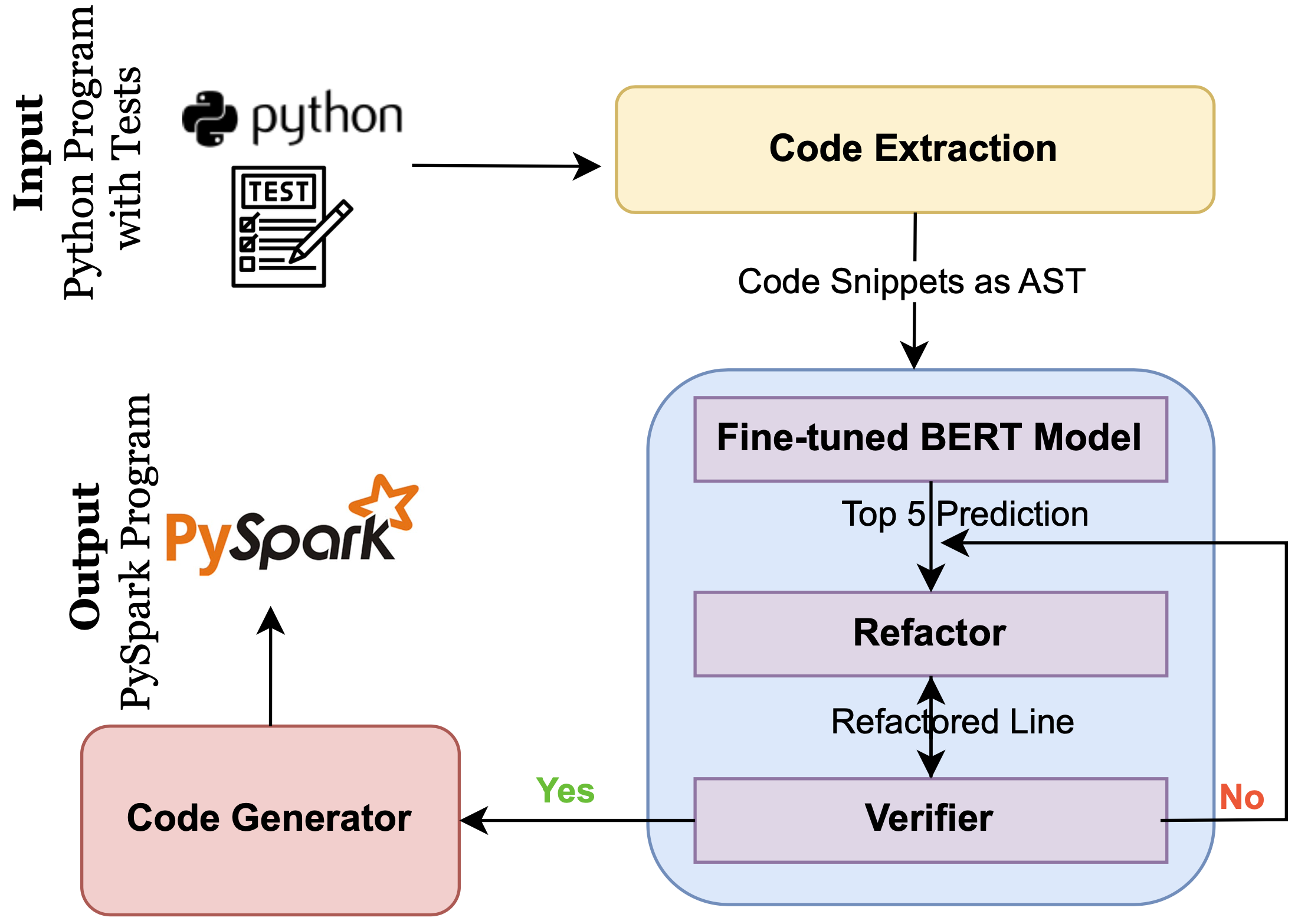}
    \caption{System Architecture of ROOP: A modular approach in translating sequential Python code into distributed PySpark applications.}
    \label{fig:architecture}
\end{figure}

The ROOP architecture methodically addresses the task of translating sequential code into a distributed application. As depicted in Fig.~\ref{fig:architecture}, the system follows a modular approach at its core, where each module focuses on a specific aspect of the translation process. From initial code parsing of the Python file to the final output, the system aims to maintain the semantics of the original code while ensuring it aligns with distributed programming principles. We explain the system's architecture components in the following sections.

The algorithm, as shown in Algorithm~\ref{alog:ROOP}, serves as the backbone for translating a sequential Python program \( P \) into its distributed form \( P' \). It initiates by parsing \( P \) using its Abstract Syntax Tree (AST) to extract relevant structures such as loops, operations, and variables. These structures are then subjected to prediction through an NLP model \( M \), thereby producing a set of refactorings, termed as \( \text{candidates} \).

For each \( \text{candidate} \) in \( \text{candidates} \), the algorithm iteratively proceeds to refactor each extracted structure. The refactored structure is tested using the \say{pytest} framework with the user-provided test program \( T \). If all tests pass successfully, the refactored structure is integrated into \( P' \). Conversely, the algorithm moves to the next candidate.

The algorithm concludes by either outputting \( P' \) as the final distributed code, provided all structures have been successfully refactored, or by raising a \say{No Translation Found} exception.

\begin{algorithm}
\caption{Sequential to Distributed Code Translation}
\label{alog:ROOP}
\begin{algorithmic}
\STATE \textbf{Input:} Sequential Program \( P \), Test Program \( T \)
\STATE \textbf{Output:} Distributed Program \( P' \)
\STATE \textbf{Parameters:} NLP Model \( M \)
 
    \STATE Parse \( P \) using AST to extract structures
    \STATE Identify and extract loops, operations, and variables
    \STATE \( \text{candidates} \leftarrow M.predict(\text{structures}) \)
    \FOR{each \( \text{candidate} \) in \( \text{candidates} \)}
        \FOR {each structure in \( \text{structures} \)}
            \STATE Refactor structure based on \( \text{candidate} \)
            \STATE Test refactored structure using `pytest` with \( T \)
            \IF {All tests are passes}
                \STATE Integrate refactored structure into \( P' \)
            \ELSE
                \STATE Continue to the next candidate
            \ENDIF
        \ENDFOR
    \ENDFOR
    \IF {All structures successfully refactored}
        \RETURN \( P' \) as final distributed program
    \ELSE
        \STATE Raise No Translation Found
    \ENDIF

\end{algorithmic}
\end{algorithm}
\subsection{Code Extraction}
Our code extraction module employs a deep traversal of Python's AST to identify and isolate significant code constructs. During this traversal, the module looks for loop nodes (for or while) and operational nodes (e.g., conditional statements, function calls, and method calls). Special attention is given to nested loops, where each inner loop is also captured as a distinct \say{Loop} data structure and linked to its parent loop for hierarchical representation.

The \say{Loop} data structure incorporates a variety of attributes, such as a unique identifier, start and end lines, and an associative list of input and output datasets. This allows for a complete understanding of the loop's role within the code. Similarly, every operation—be it a simple arithmetic operation, a conditional check, or a dataset manipulation—is encapsulated within an \say{Operation} data structure. This structure stores pertinent details like the involved variables, the line number, and the operation type.

To ensure exhaustive extraction, the AST nodes are traversed recursively, enabling the identification of operations even within nested conditions and loops. Each \say{Loop} and \say{Operation} data structure is then populated dynamically as the AST is traversed, thereby constructing an organized and detailed snapshot of the code's logic and functionality.

Table~\ref{tab:extractions} provides a concrete example of how our code extraction module operates on a Python function designed to filter even numbers from a list. It showcases the original Python code alongside the loop information extracted and formatted in a structured JSON-like representation. This extracted information highlights key attributes such as the loop's unique identifier, start, and end lines, involved datasets, and operations—attributes that are dynamically populated by traversing the AST.
\begin{table}[t!]
\caption{Loop information extracted from a Python function represented in JSON-like format.}
\centering
\label{tab:extractions}
\begin{tabular}{|l|l|}
\hline
\textbf{Code Snippet} & \textbf{Extracted Loop Information} \\
\hline
\tiny
\begin{lstlisting}[style=pythonstyle]

def even_filter(numbers):
    evens = []
    for num in numbers:
        if num % 2 == 0:
            evens.append(num)
    return evens
\end{lstlisting} 
&
\tiny\begin{lstlisting}[style=pythonstyle]
{
"Loop ID": 1,
"Start Line": 3,
"End Line": 6,
"Is Nested": "No",
"Datasets": {
"Input": "numbers",
"Output": "evens"
},
"Operations": [
{
  "Type": "Conditional",
  "Expression": "num % 2 == 0"
},
{
"Type": "Method Call",
"Expression": "evens.append(num)"
}]}
\end{lstlisting} \\
\hline
\end{tabular}
\end{table}

\subsection{ Fined-tuned BERT Model}
A key component of the ROOP system is its Fine-Tuned BERT model for code translation. The model architecture is derived from the pre-trained BERT-base model and is specialized for our translation task. The BERT Classifier class serves as the backbone for our translation model, incorporating both the BERT tokenizer and the TensorFlow-based sequence classification model. The classifier is initialized with a pre-trained model stored locally, and it is fine-tuned to classify 111 different labels which correspond to the PySpark API calls~\cite{tyro3}.

To handle the sequence inputs, we employed BERT's tokenizer with a maximum sequence length of 256. The input code snippets are tokenized and fed into the model, which then returns a set of logits. These logits are transformed into probabilities using a softmax layer. The model predicts the top 5 candidates or PySpark API calls that are most likely to correspond to the input code snippet. The top predicted labels are inverse-transformed to get their corresponding PySpark API calls. Additionally, the class includes methods for making single predictions, offering flexibility based on the task at hand. This fine-tuned BERT model serves as a powerful mechanism for highly accurate code translation, contributing to the overall robustness and efficiency of the ROOP system. 

Table~\ref{tab:ModelPrediction} provides an illustrative example of how the fine-tuned BERT model performs in predicting distributed operations for given Python loop code. The table shows three different Python loops alongside the top 5 PySpark API calls predicted by the model. For instance, the loop that filters even numbers from a list has candidates like 'flatMap(),count()', 'filter()', and 'sortBy()', among others. These predicted candidates demonstrate the model's capability to suggest a variety of PySpark methods that could be used to achieve the same functionality as the original Python code but in a distributed environment.

Similarly, for a loop that converts strings to lowercase and accumulates them, the model predicts operations like 'filter(),sum()' and 'reduce(),collect()', revealing its understanding of both data filtering and aggregation tasks in a distributed setting. Another loop, which flattens a list of lists, has predictions like 'flatMap(),distinct(),count()' and 'flatMap(),sum()', suggesting that the model can identify complex nested operations that can be parallelized.

\begin{table}[t!]
\caption{Top $5$ distributed operations predictions by the fine-tuned BERT model.}
\label{tab:ModelPrediction}
\centering
\footnotesize
\begin{tabular}{|l|l|}
\hline
\textbf{Original Loop Code} & \textbf{Top 5 Predicted Distributed Operations} \\
\hline
\begin{lstlisting}[language=Python]
for num in numbers:
    if num % 2 == 0:
        evens.append(num)
\end{lstlisting}
&
\begin{lstlisting}[language=Python]
`flatMap(),count()', 
`take()', 
`sortBy()',
`map(),distinct()', 
`filter()'
\end{lstlisting} \\ 
\hline
\begin{lstlisting}[language=Python]
for str in strings:
        lower = str.lower()
        result += lower
\end{lstlisting}
&
\begin{lstlisting}[language=Python]
`filter(),sum()',
`reduce(),collect()', 
`sum()',
`map(),reduce()',
`reduce()'
\end{lstlisting} \\ 
\hline
\end{tabular}
\end{table}
\subsection{Refactor}
The Refactor module operates together with the Code Extraction module, forming a cohesive solution for transforming Python code into PySpark snippets. This module accommodates a wide array of data manipulation operations, including but not limited to \textit{map}, \textit{filter}, \textit{reduce}, \textit{join}, \textit{union}, \textit{sort}, \textit{groupBy}, \textit{flatMap}, \textit{sum}, and \textit{count}.

Each data manipulation operation, be it \textit{map}, \textit{filter}, \textit{reduce}, or \textit{join}, has  a method within the Refactor module to handle its specific requirements. For example, the \textit{map\_operation} method corresponds to the \textit{map} function and takes parameters such as the source dataset, the details of the operation to be executed, and the resulting dataset. 

Additionally, each PySpark API call is associated with a specialized method tailored to its parameter requirements. For example, operations involving multiple datasets, such as \textit{join} or \textit{union}, accept additional parameters to specify the secondary datasets. This adaptability ensures that the module can handle a variety of data manipulation tasks. On the other hand, simpler methods like \textit{sum} and \textit{count} do not require any extra parameters, making them straightforward to use. Overall, this detailed approach allows the Refactor module to offer a rich set of functionalities through distinct methods.

Diving into the specifics of how individual loop fragments are transformed into PySpark code snippets, the refactor module relies on the information extracted by the Code Extraction module. This includes details like the list of datasets, the operations performed within the loop, and any conditions as shown in Table~\ref{tab:extractions}. Once the BERT model provides its predictions as candidates, the Refactor module gets to work, iteratively applying the refactoring functions for each API call suggested by the model.

For instance, consider an operation like \textit{lower = str.lower()}. It is first wrapped into a lambda function as \textit{lambda lower: str.lower()}. This lambda function is then wrapped by the \textit{map} method, as predicted by the BERT model, resulting in \textit{map(lambda lower: str.lower())}.

Similarly, for filter operations that primarily work on conditions, a condition like \textit{num \% 2 == 0} would be encapsulated within a lambda function as \textit{lambda num: num \% 2 == 0}. This is then further wrapped by the \textit{filter} method to form \textit{filter(lambda num: num \% 2 == 0)}.

Importantly, each method accepts two key parameters: the primary dataset and the result dataset or variable. The refactored code is first concatenated with its primary dataset and then assigned to the result dataset. For example, the refactored code \textit{map(lambda lower: str.lower())} would concatenate with its primary dataset, \textit{strings\_rdd}, and then be assigned to the result dataset \textit{strings\_rdd} to form \textit{strings\_rdd = strings\_rdd.map(lambda lower: str.lower())}. This approach is systematically applied to each prediction, transforming them efficiently and coherently into PySpark API calls. Similarly for the filter operations, it does the same and returns \textit{evens = numbers\_rdd.filter(lambda num: num \% 2 == 0)}

The Table~\ref{tab:refactor} serves as an illustration of the refactoring module's capabilities. On the left, we have a conventional loop in Python and on the right, the refactored code accomplishes the same task but uses the distributed PySpark API calls.

\begin{table}[ht]
\caption{Example illustrating the transformation of original loop code into refactored code.}
\label{tab:refactor}
\centering
\footnotesize
\begin{tabular}{|l|l|}
\hline
\textbf{Original Loop Code} & \textbf{Refactored Code} \\
\hline
\begin{lstlisting}[style=pythonstyle]
for num in numbers:
    if num % 2 == 0:
        evens.append(num)
\end{lstlisting}
&
\begin{lstlisting}
evens = numbers_rdd
  .filter(
    lambda num: num % 2 == 0
  )
\end{lstlisting} \\ 
\hline
\begin{lstlisting}[style=pythonstyle]
 for sublist in list_of_lists:
        for item in sublist:
            result.append(item)
\end{lstlisting}
&
\begin{lstlisting}
list_of_lists_rdd
  .flatMap(lambda x: x)
  .collect()
\end{lstlisting} \\ 
\hline
\end{tabular}

\end{table}

\subsection{Verifier}

In ROOP's pipeline, verifying the correctness of the generated PySpark code is a crucial step. While the theoretical problem of program equivalence is undecidable~\cite{boyer1984mechanical}, ROOP utilizes a practical approach through unit testing to confirm that the refactored code is equivalent to the original Python code. 

The Verifier module, central to this step, incorporates several methods that together assure code equivalence. Initially, the user needs to provide unit tests for the original Python code, which serves as the ground truth. ROOP assumes that the original code passes all these tests.

With the assistance of the Python package \textit{pytest} and a specialized version customized for Spark, known as \textit{pytest-spark}, the \textit{run\_pytest\_isolated} method executes these unit tests on the generated PySpark code without requiring any alterations. The module encapsulates SparkContext within each function, thereby eliminating the issue of multiple connections. Additionally, ROOP sidesteps the need to import any libraries during the testing phase. This is possible because \textit{pytest} can be invoked via Python's \textit{subprocess} module, which also captures standard output and error streams for further analysis.

To check for successful verification, the method \textit{parse\_test\_xml} reads the \textit{pytest} output logged in XML format. Any failing test case along with its failure message is collected for debugging purposes. If all tests pass and the standard error stream is empty, the generated code is considered verified as it passes all the unit tests and is equivalent to the original Python code.

This verification process is looped through for each generated prediction. The code is first written to a temporary Python file which is generated with the help of the Code Generation module (described in Section~\ref{sec:code_gen}), after which the \textit{run\_pytest\_isolated} method is invoked to perform the testing. If all tests pass, the prediction is  correct, and the loop breaks, confirming that the refactored code performs identically to the original code. 

This comprehensive approach ensures that the final program, including the refactored code, is fully functional and ready for execution.

\subsection{Code Generator}\label{sec:code_gen}

Upon invocation, the function initiates the PySpark environment by generating a function, \textit{get\_or\_create\_spark\_context}, which ensures that a Spark context is either retrieved or created. This function sets the application name and runs the Spark master locally. This initialization ensures that each function within the code has a unique Spark context.

 The previous module Code Extraction has already stored loop information for us. It has already mapped a loop with its starting line number and end line number. The Code Generator iterates through each line of the original code to find the corresponding loops. Upon finding a loop, it adds the PySpark initialization code along with spark context and then replaces the original loop with the refactored PySpark code given by the Verifier module after successful validation. The function pays special attention to the indentation to ensure that the refactored code aligns correctly with the original Python code.

Additionally, for each dataset used in the loop, the function parallelizes it using SparkContext that it initialized earlier, thereby converting it into an Resilient Distributed Dataset (RDD). If a loop involves a \textit{flatMap} operation, the loop skips parallelization for the remaining datasets to avoid redundancy since the secondary dataset in flatmap is not required to be parallelized. However, for APIs like \textit{join} and \textit{union}, it parallelizes both primary and secondary datasets. 

After inserting the refactored code, the function adds a command to  stop the spark context. This ensures that no unnecessary resources are consumed. This is done at the end of the loop, which again is extracted by the Code Extraction module earlier as the end line number. 

Finally, the module stitches together all these components, preserving the original code structure where no changes are needed, and returns the executable PySpark code.

\section{Experimental Setup and Metrics}

\subsection{Setup}

The experiments were conducted in a controlled environment to ensure consistent and reliable results.

\textbf{For Fine-tuned BERT Model:}
An Intel Xeon Gold 6258R CPU running at 2.00GHz with 6 cores, an NVIDIA Tesla V100 GPU with 15GB of memory, 32GB of DDR4 RAM, and a 256GB NVMe SSD were all used as the hardware and software setup for training and testing the model. The deep learning library used was TensorFlow 2.12.0, and the computing environment was Python 3.10.12. Tokenization also made use of the BERT Tokenizer from HuggingFace's Transformers library.

\textbf{For ROOP Synthesis}
An Intel Core i5 8250U CPU running at 1.80GHz with 8 cores, 8 GB of DDR4 RAM, and a 256GB NVMe SSD were the main parts of the hardware setup used for `ROOP'. Python 3.11.4 was used for the development and testing of ROOP. For testing, we used the Python Package PyTest. 

\textbf{Test Suites:} 
In our experiments, we utilized a diverse array of Python programs to ensure a comprehensive evaluation. The dataset features $14$ distinct programs, which contain a total of $26$ loop or iterative fragments. These fragments range from simple loop constructs to complex nested structures. The programs are organized into three distinct test suites, each targeting varying levels of computational complexity:

\begin{itemize}
  \item \textbf{Simple Operations:} Programs using single or sequentially chained PySpark API calls like map, reduce, and filter within a single loop.
  
  \item \textbf{Nested Operations:} Programs with nested loops or loops collaborating across multiple datasets, involving operations such as union, flatmap, and join.
  
  \item \textbf{Complex Operations:} Programs with higher computational complexity, featuring multiple nested or sequential loops, each encapsulating more than two PySpark API operations, often chained or interdependent.
\end{itemize}

We assessed the performance of our system using these principal metrics, each of which provides a different perspective on the system's overall efficacy:

\begin{itemize}
    \item \textbf{Translation Accuracy:} This metric gauges the system's ability to correctly translate Python programs into PySpark. A high translation accuracy ensures that the converted programs maintain the original program's logic and functionality.
    
    \item \textbf{Synthesis Speed:} This metric evaluates the amount of time taken by ROOP to generate the PySpark translation from a given Python program. A faster synthesis speed is advantageous for real-time or near-real-time applications. 
\end{itemize}

\section{Results and Discussion}
\begin{figure*}[b!]
    \centering
     \includegraphics[width=\textwidth]{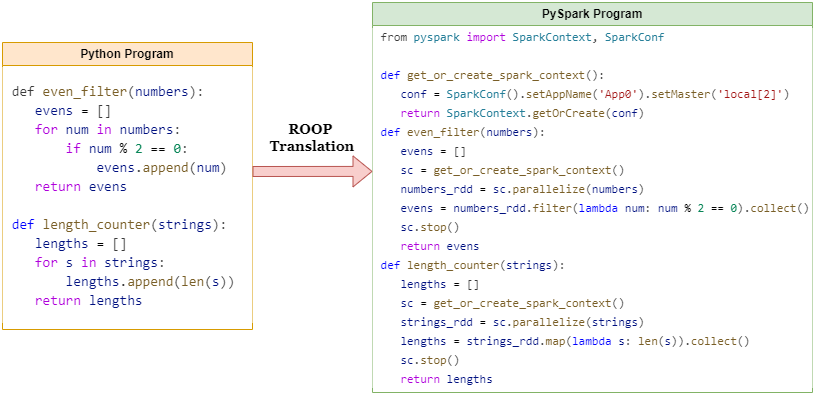}
    \caption{Translation of ROOP from Python to executable PySpark}
    \label{fig:translation}
\end{figure*}

Fig.~\ref{fig:translation} is an example of ROOP's capability to convert standard Python functions into their PySpark equivalents. For instance, the Python function \textit{even\_filter} uses a traditional for-loop to filter out even numbers from a list. In its PySpark counterpart, the function leverages Spark's Resilient Distributed Datasets (RDD) and the \textit{filter} transformation to achieve the same outcome, but in a distributed computing environment. Similarly, the \textit{length\_counter} function in Python, which calculates the lengths of given strings, is translated to use PySpark's \textit{map} transformation. Both translated functions also manage the creation and termination of a Spark context, showcasing ROOP's ability to seamlessly integrate resource management within the translated code.

The \textit{get\_or\_create\_spark\_context} function serves as a centralized manager for SparkContext, ensuring efficient and safe usage within the application. It utilizes Spark's getOrCreate method to either retrieve an existing SparkContext or create a new one, thereby avoiding the pitfalls of multiple initialization. This singleton approach simplifies resource management and allows for consistent configuration across the application. 

Through these example functions, Fig.~\ref{fig:translation} highlights the adaptability and efficacy of ROOP in enabling distributed computing capabilities by translating Python logic into PySpark. Such translations are invaluable for tasks that are data-intensive and require scalability across multiple nodes in a Spark cluster.

\subsection{Translation Accuracy}
Table~\ref{tab:translation_accuracy} summarizes ROOP's effectiveness in converting Python programs to their PySpark equivalents. The dataset is divided into three separate test suites for a more in-depth analysis: Simple Operations, Nested Loops, and Complex Programs. These suites consist of 14 programs with 26 loop or iterative fragments in total. Examining the table closely indicates a remarkable translation accuracy rate. All but one loop segments across all suites could be translated with success by ROOP. 

The incorrect translation that we encountered during our evaluation was the program that sums up all the even numbers in a given list. The reason for this failure is the model's inability to accurately predict the appropriate PySpark API calls that correspond to the original Python code. Specifically, the model struggled to recognize that the Python for loop and the conditional if statement could be translated into a PySpark \textit{filter} followed by a \textit{reduce} operation. As part of our future work, we plan to enhance the model's capabilities by training it on a more diverse set of code snippets. This will help us to improve the accuracy of API call predictions, thereby reducing such failures

To sum it up, the high rate of successful translations across diverse operational complexities validates the reliability and generalizability of ROOP's translation mechanism. While one instance of translation failure was observed, this doesn't undermine ROOP's overall effectiveness; rather, it identifies an avenue for future improvement. This insight informs our future work, where we plan to enhance the model by training it on a more diverse set of code snippets, thereby improving its  accuracy and adaptability across different programming scenarios.

\begin{table}[ht]
\caption{Translation Accuracy Results}
\label{tab:translation_accuracy}
\centering
\begin{tabular}{|c|l|l|l|l|}
\hline
Test  & Total  & \# of Loop  & Successful \\
Suites& Programs& Fragments&Translations \\
\hline
Simple Operations     & 6   & 7      &  6     \\
Nested Operations     & 5   & 10            & 10 \\
Complex Operations & 3   & 9            & 9 \\
Overall          & 14   & 26            & 25 \\
\hline
\end{tabular}

\end{table}


\subsection{Synthesis Speed}

\begin{table*}[t!]
\caption{Execution Time Metrics Across Test Suites}
\label{tab:sysnthesisspeed}
\centering
\begin{tabular}{|l|l|r|r|r|r|}
\hline
     Folder             & Program                   &   Avg &   Min &   Max &   Std \\
\hline
  complex\_operations & flatmap\_distinct\_count.py &   152.16  &   129.89  &   248.46  &  35.71  \\
  complex\_operations & flatmap\_filter\_sort.py    &   108.16 &    86.29 &   157.71  &  21.57  \\
   complex\_operations & map\_distinct\_sort.py      &   103.61  &    86.76 &   116.49  &   7.41 \\
  nested\_operations  & flatmap.py                &   121.45  &    98.31 &   156.3  &  17.96  \\
  nested\_operations  & flatmap\_count.py          &    94.28 &    76.91 &   129.76  &  17.18    \\
   nested\_operations  & join.py                   &   147.27  &   105.00  &   205.21  &  32.34  \\
   nested\_operations  & union.py                  &    85.61 &    66.91  &   111.02 &  15.39  \\
   nested\_operations  & union\_count.py            &    57.45 &    46.43 &    75.70 &   8.95 \\
  simple\_operations  & filter\_count.py           &    44.02 &    35.11 &    65.10 &   9.37 \\
  simple\_operations  & filter\_reduce.py          &    82.18 &    70.87&   111.95  &  12.15   \\
 simple\_operations  & map\_reduce.py             &    67.64 &    58.13   &    85.67 &   8.06 \\
  simple\_operations  & map\_sum.py                &    84.73 &    72.37  &   106.78  &  10.66  \\
  simple\_operations  & multiple\_loop.py          &    53.92 &    41.97 &    94.36 &  15.5898  \\
  simple\_operations  & reduce.py                 &    53.10 &    45.37 &    63.67  &   7.66 \\
\hline
\end{tabular}



\end{table*}
To assess the efficiency of the ROOP system in translating Python programs into their PySpark equivalents, we present a comprehensive set of metrics in Table~\ref{tab:sysnthesisspeed}.  The programs are categorized into three distinct test suites: Simple Operations, Nested Operations, and Complex Operations, to provide a more nuanced understanding of performance. For statistical robustness, each program was subjected to 10 translation iterations, with metrics like mean, minimum, maximum, and standard deviation of the execution times calculated.

Upon close examination of Table~\ref{tab:sysnthesisspeed}, it becomes evident that program complexity is a significant factor influencing translation time. Specifically, the \textit{Complex Operations} suite exhibits longer execution times, peaking at 248.46 seconds for the program \textit{flatmap\_distinct\_count.py}. 

Conversely, the \textit{Simple Operations} suite demonstrates efficiency, with the \textit{filter\_count.py} program requiring a maximum of only 65.10 seconds for translation.  It is noteworthy that the prediction phase of the BERT model runs within 7 to 10 seconds on average. However, because a new test environment needs to be set up for each prediction, the verification process adds a significant amount of time overhead. In conclusion, improving the correctness of the BERT model to reduce verification occurrences and maybe utilizing distributed computing resources to accelerate the verification process are two possible ways to increase performance efficiency.

\section{Conclusion and Future Work}
This work introduces ROOP, an innovative tool that uses a BERT-based NLP model to swiftly and accurately translate Python programs into PySpark counterparts. Our experimental evaluation, which included 14 different Python programs and 26 loop pieces, highlights the tool's robustness and versatility. ROOP translated 25 of 26 loop pieces effectively, indicating great accuracy across a wide variety of complexities. Furthermore, our metrics reveal that for simpler procedures, ROOP can finish translations in as little as 44 seconds, showcasing its efficiency.

However, like with any initial work, there are opportunities for improvement and future research. The translation time for more complex operations, which can take several minutes, is one drawback. Future work might focus on optimizing the BERT model and the process of verification to further minimize these times. One such path could be the implementation of parallel or distributed verification. Furthermore, while ROOP has great translation accuracy, it did fall short in one case. Investigating the reasons behind this and tweaking the model accordingly may help it perform more effectively.

Another exciting direction would be to extend ROOP's capabilities to other distributed computing frameworks beyond PySpark. Given the tool's underlying architecture, adapting it to other languages or frameworks is a feasible and intriguing avenue for future research.

In summary, ROOP shows great promise as a tool for simplifying the transition from traditional to distributed programming. Its high accuracy and speed can significantly streamline the software development process, making it a valuable asset for both researchers and practitioners in the field of distributed computing.

\begin{credits}
\subsubsection{\discintname}
The authors have no competing interests to declare that are relevant to the content of this article.
\end{credits}

\bibliographystyle{splncs04}
\bibliography{ref}

\end{document}